\documentclass{emulateapj}
\usepackage{apjfonts}
\usepackage{graphicx}
\usepackage[svgnames]{xcolor}

\newcommand{\hubble}{\textit{Hubble Space Telescope}}
\newcommand{\hst}{\textit{HST}}
\newcommand{\chandra}{\textit{Chandra}}
\newcommand{\cxo}{\textit{Chandra X-ray Observatory}}

\newcommand{\ergsec}{\mbox{erg s$^{-1}$}}
\newcommand{\err}[2]{\ensuremath{^{_{+#1}}_{^{-#2}}}}
\newcommand{\ee}[2]{\ensuremath{{#1}\times10^{#2}}}

\newcommand{\cv}{\textcolor{blue}{$\blacktriangle$}}
\newcommand{\qlmxb}{\textcolor{red}{\scriptsize $\blacksquare$}}
\newcommand{\psrm}{\textcolor{IndianRed}{$\blacktriangledown$}}
\newcommand{\psr}{\textcolor{DarkGreen}{$\blacktriangledown$}}
\newcommand{\ab}{\textcolor{Goldenrod}{$\bigstar$}}
\newcommand{\fpsr}{\textcolor{BlueViolet}{\Large $\bullet$}}
\newcommand{\fbrst}{\textcolor{Peru}{\Large $\bullet$}}

\begin{document}
%%%%%%%%%%%%%%%%%%%%%%%%%%%%%%%%%%%%%%%%%%%%%%
\shorttitle{Dynamical Formation of Globular Cluster CVs}
\shortauthors{Pooley \& Hut}
\slugcomment{submitted to ApJL}
%%%%%%%%%%%%%%%%%%%%%%%%%%%%%%%%%%%%%%%%%%%%%%
\title{Dynamical Formation of Close Binaries in Globular Clusters II: Cataclysmic Variables}

\author{David Pooley\altaffilmark{$\star$}}
\affil{University of California at Berkeley\\ Astronomy Department, 601 Campbell Hall, Berkeley, CA 94720}
\email{dave@astron.berkeley.edu}

\and 
\author{Piet Hut} 
\affil{Institute for Advanced Study\\ Einstein Drive, Princeton, NJ 08540}
\email{piet@ias.edu}

\altaffiltext{$\star$}{Chandra Fellow}
%%%%%%%%%%%%%%%%%%%%%%%%%%%%%%%%%%%%%%%%%%%%%%
\begin{abstract}

We answer the long-standing question of which production mechanism is responsible for the cataclysmic variables (CVs) in globular clusters.  Arguments have been given that range from mostly primordial presence to a significant contribution of later dynamical formation in close stellar encounters.  We conclude, based on a thorough analysis of a homogeneous {\it Chandra} data set, that the majority of CVs in globulars has a dynamical origin.

\end{abstract}
%%%%%%%%%%%%%%%%%%%%%%%%%%%%%%%%%%%%%%%%%%%%%%

\section{Introduction}

In the 1970s, it was discovered that the number of outbursting low-mass X-ray binaries (LMXBs) per unit mass was orders of magnitude higher in globular clusters than the rest of the Milky Way \citep{1975ApJ...199L.143C, 1975Natur.253..698K}, and theoretically good reasons were quickly put forward to
explain how LMXBs could be formed efficiently through encounters in
globulars \citep{1975MNRAS.172P..15F, 1975MNRAS.173..729H, 1975AJ.....80..809H, 1976MNRAS.175P...1H, 1975A&A....44..227S}.  Taking an LMXB and replacing the neutron star by a white dwarf, we have a cataclysmic variable (CV).
Whether globulars would also enhance CV formation through encounters
was less clear.  \citet{1983Natur.301..587H} 
concluded that there still would be a contribution, but a much smaller
one, and a detailed study by \citet{1994ApJ...423..274D} came to the same conclusion.

There have been two main difficulties in testing this prediction.  Until recently, globular cluster CVs have proven rather elusive, and only a handful were known in the entire Galactic globular cluster population.  However, with high resolution data from the \cxo\ and the \hubble, we are now finding dozens of CVs per cluster \citep[e.g.][]{2001Sci...292.2290G,2002ApJ...569..405P}.  Another difficulty has been that the density of CVs in the field is not well known, so there is no basis for comparison.   Recently, \citet{2005ApJ...628..395T} have come up with a theoretical estimate for the number of CVs per unit mass of an old stellar population, and they concluded that the number of X-ray sources in 47 Tuc is compatible with their estimate, showing no overabundance relative to the field.  This would mean that encounters have little to no effect on CV production, in contrast to the outbursting LMXBs.

On the basis of \chandra\ observations of a dozen globular clusters, \citet[][paper I]{2003ApJ...591L.131P} found that the number of X-ray sources in a cluster scales with the encounter frequency ($\Gamma$) of the cluster significantly better than it scales with the mass of the cluster.  This was the first clear proof that the population of X-ray sources in a globular cluster is largely dynamically formed.

Many of the X-ray sources in a number of clusters have been identified, and it has become clear that this is a heterogeneous population, including LMXBs (in quiescence for the clusters under consideration here), CVs, chromospherically active main-sequence binaries, and millisecond pulsars.  It was recognized that the quiescent LMXBs (qLMXBs) could be isolated from the rest of the population for the most part, and a number of authors noted the near-linear relationship between the number of qLMXBs and $\Gamma$ \citep{2003ApJ...591L.131P, 2003A&A...403L..11G, 2003ApJ...598..501H}.  Because this relationship was based on the number of total LMXBs (not just the outbursting ones), this firmly established that a higher rate of encounters resulted in a higher number of LMXBs, not just a higher fraction of outbursting systems, as could have been argued for the long-standing result from the 1970s.

Having separated the qLMXB population from the general population, it is easily seen that qLMXBs are the minority subpopulation.  Because the correlation of \citet{2003ApJ...591L.131P} was based on the entire population, the implication is that, presumably, the other subpopulations must also receive some contribution to their formation from dynamical encounters.  In this paper, we address the CV subpopulation.  We sketch our approach in \S\ref{sec:method} and \S\ref{sec:sample}, analyse it critically in \S\ref{sec:analysis}, and sum up in \S\ref{sec:summary}.

\section{A Clear Encounter Signal}
\label{sec:method}
The main difficulty in our previous paper was the strong correlation between a cluster's mass ($M$) and its encounter frequency $\Gamma$, defined as $\int_0^{r_h} (\rho^2/v)\, 4 \pi r^2 dr$, where $r_h$ is the half-mass radius, $\rho$ is the stellar density as a function of radius, and $v$ is the velocity dispersion as a function of radius.  This integral is often approximated as ${\rho_0}^2 {r_c}^3 / v_0$, where $r_c$ is the core radius and the subscript zero refers to central values; this expression simplifies to ${\rho_0}^{1.5}{r_c}^2$ for a virialized system.  By and large, clusters with a larger mass have a larger integrated collision number, and vice versa.  Performing a correlation test between $M$ and $\Gamma$ for the 140 globular clusters in the \citet{1996AJ....112.1487H} catalog, we find a Spearman coefficient of 0.72 with a chance correlation probability of $\sim$10$^{-23}$.  However, there is a very large scatter in the correlation.  At a given $M$, there are orders of magnitude variation in $\Gamma$; the best fit line between $\log M$ and $\log \Gamma$ has a mean absolute deviation of 0.69. 

In this paper, we alleviate the problem of the $\Gamma$--$M$ correlation by working in specific units.  We define $\gamma = \Gamma/M$ as the specific encounter frequency and $n_x = N_x/M$ as the specific number of sources of population $x$.  The specific encounter frequency $\gamma$ is a measure of the chance that {\it a particular} star in a globular cluster undergoes an encounter, in contrast to $\Gamma$, which is the chance that {\it any} star in a cluster has an encounter.  We estimate the mass of the cluster as $M/M_\sun = 3\times10^{0.4(M_{V,\sun}-M_V)}$ using the cluster absolute magnitudes from the 2003 version of the catalog of \citet{1996AJ....112.1487H}, and we have used units of $10^6 M_\sun$ to form our specific units.

With these units, we can perform a very simple test.  If population $x$ is primordial, then $n_x$ should be roughly the same for all clusters (where we neglect variations in the initial mass function (IMF), etc.; see \S\ref{sec:analysis}).  This is easily checked by fitting for a constant, $n_x = C_x$.  Should this give a statistically unacceptable fit, we conclude that population $x$ is not simply primordial.  To  test for the effects of encounters, we can take the next step in complexity and assume some primordial contribution and some dynamical contribution: $n_x = C_x + A_x\gamma^{\alpha_x}$.

\begin{deluxetable}{cccl}
\tablewidth{0pt}
\tablecaption{Results of fitting $n_x = C_x$\label{tab:const}}
\tablehead{\colhead{$x$} & \colhead{$C_x$} & \colhead{$\chi^2/\mathrm{dof}$} & \colhead{$Q$ value}} \\
\startdata
I & 1.29 & 27.8/5 & \ee{3.91}{-5} \\          %equivalent sigma = 4.1
II & 3.99 & 85.5/22 & \ee{1.97}{-9} \\        %equivalent sigma = 6.0
III & 16.8 & 69.1/18 & \ee{6.41}{-8}\\        %equivalent sigma = 5.4
I+II+III & 22.0 & 120.7/18 & \ee{1.12}{-17}   %equivalent sigma = 8.6
\enddata
\end{deluxetable}

\section{A (Relatively) Clean Sample of CVs}
\label{sec:sample}
Determining how many of the \chandra\ X-ray sources in a globular cluster are CVs is difficult because most of the sources are discovered with only tens of X-ray counts.  It is only through a comparison with \hst\ data to look for blue or H$\alpha$ bright sources within the X-ray error circles that most CVs are identified (this requires the matching of the \chandra\ and \hst\ frames to sub-arcsecond precision, which is nontrivial).  This is a time-consuming process and has only been accomplished for a large number of sources in just a few clusters.  Here, we outline a method to obtain a relatively clean sample of CVs in each cluster on the basis of the X-ray data alone.

\begin{figure*}
\centering
\includegraphics[width=0.85\textwidth]{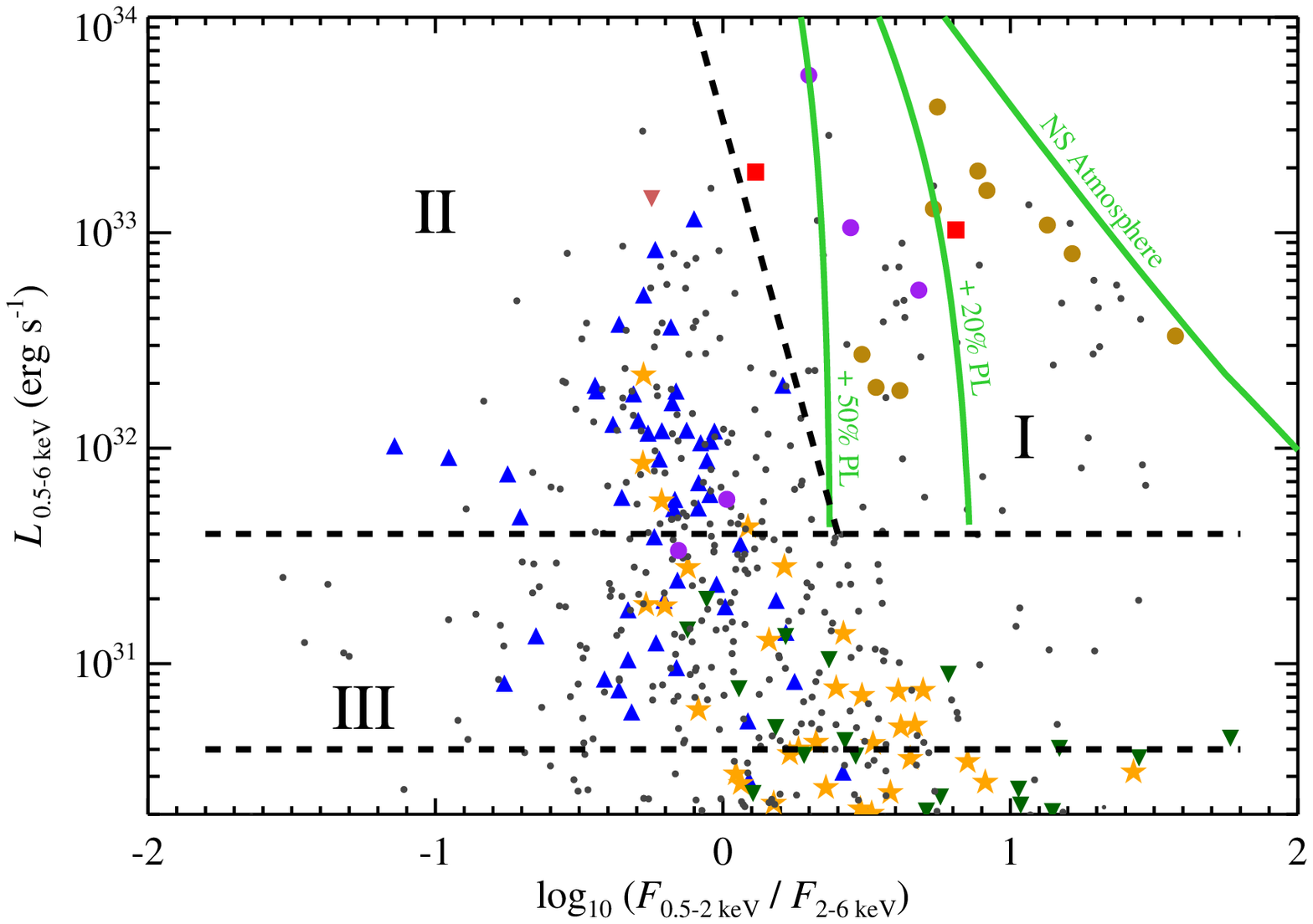}
\caption{The \chandra\ color-magnitude diagram of 23 globular clusters, corrected for absorption. Above \ee{4}{30}~\ergsec (\ee{4}{31}~\ergsec), there are $\sim$500 ($\sim$200) sources in this diagram, about 100 (15) of which are expected to be background sources. Not all clusters have been observed to the same limiting luminosity.  Securely identified cluster sources are denoted by \qlmxb\ for qLMXBs, \cv\ for CVs, \ab\ for active main-sequence binaries, \psr\ for millisecond pulsars, and \psrm\ for the pulsar in M28. In addition, we show the absorption-corrected locations of numerous field qLMXBs: \fpsr\ for pulsars and \fbrst\ for bursters. Tracks for NS atmosphere spectral models (plus some power law contributions) are also plotted; these spectral models have been used to successfully fit many qLMXBs.}
\label{fig:cmd}
\end{figure*}

In Fig.~\ref{fig:cmd}, we show an X-ray color-magnitude diagram for the X-ray sources within the half-mass radii of 23 globular clusters\addtocounter{footnote}{-1}\footnote{NGCs 104 (47 Tuc), 288, 362, 5139 ($\omega$ Cen), 5272 (M3), 5904 (M5), 6093 (M80), 6121 (M4), 6205 (M13), 6218 (M12), 6266 (M62), 6341 (M92), 6366, 6397,  6440, 6541, 6544, 6626 (M28), 6656 (M22), 6752, 6809 (M55), 7099 (M30), and Terzan 5}.  In total, $\sim$600 sources are detected in these clusters, about 500 of which are represented in the figure (the rest are at lower $L_\mathrm{x}$); we expect about 130 of these to be unrelated background sources based on the $\log N - \log S$ relationship of \citet{2001ApJ...551..624G}.  We denote those sources whose nature has been identified\footnote{The identifications come from 47 Tuc \citep{2001Sci...292.2290G,2005ApJ...625..796H}, NGC 288 \citep{2006astro.ph..3374K}, NGC 6093 \citep[M80;][]{2003ApJ...598..516H}, NGC 6121 \citep[M4;][]{2004ApJ...609..755B}, NGC 6397 \citep{2001ApJ...563L..53G}, NGC 6440 \citep{2001ApJ...563L..41I}, NGC 6626 \citep[M28;][]{2003ApJ...594..798B}, NGC 6752 \citep{2002ApJ...569..405P}, and Terzan 5 \citep{2005ApJ...618..883W}.} by various symbols.    We also plot the locations of a number of field neutron star LMXBs in quiescence.  Further, we show the tracks of some spectral models that have been successfully used to describe the spectra of quiescent neutron star LMXBs; these models consist of a thermal, blackbody-like component from the neutron star surface propogated through a Hydgrogen atmosphere \citep{1996A&A...315..141Z} plus some contribution (up to about 50\%) from a powerlaw component, whose origin is unknown.

Based on observational considerations and the theoretical spectral models, we have divided the X-ray sources in globular clusters into three populations, creatively denoted I, II, and III.  Population I includes the two secure globular cluster qLMXBs (and most of the other known quiescent LMXBs from the field) and relatively little else.  Population II includes the rest of the sources down to a luminosity limit of $4\times10^{31}$~\ergsec\ (22 globulars have been observed to this limit), and population III goes a factor of ten lower to  $4\times10^{30}$~\ergsec\ (18 globulars have been observed to this limit).  In population II, there have been a number of source identifications.  Although the constituents are mixed, the strong majority component appears to be CVs.  This makes sense observationally, since few qLMXBs are as spectrally hard as most CVs, and the chromospherically active main-sequence binaries rarely reach these luminosity levels \citep{1993ApJS...86..599D,1997ApJ...478..358D}.   There is little we can say about population III at this point except that it likely includes contributions from all known subpopulations.  Differences in the X-ray luminosity functions from cluster to cluster \citep{2002ApJ...573..184P} may suggest that the makeup of this population varies from cluster to cluster.

\section{Analysis}
\label{sec:analysis}
\begin{figure*}                                                                 
\centering                                                                 
\includegraphics[width=0.47\textwidth]{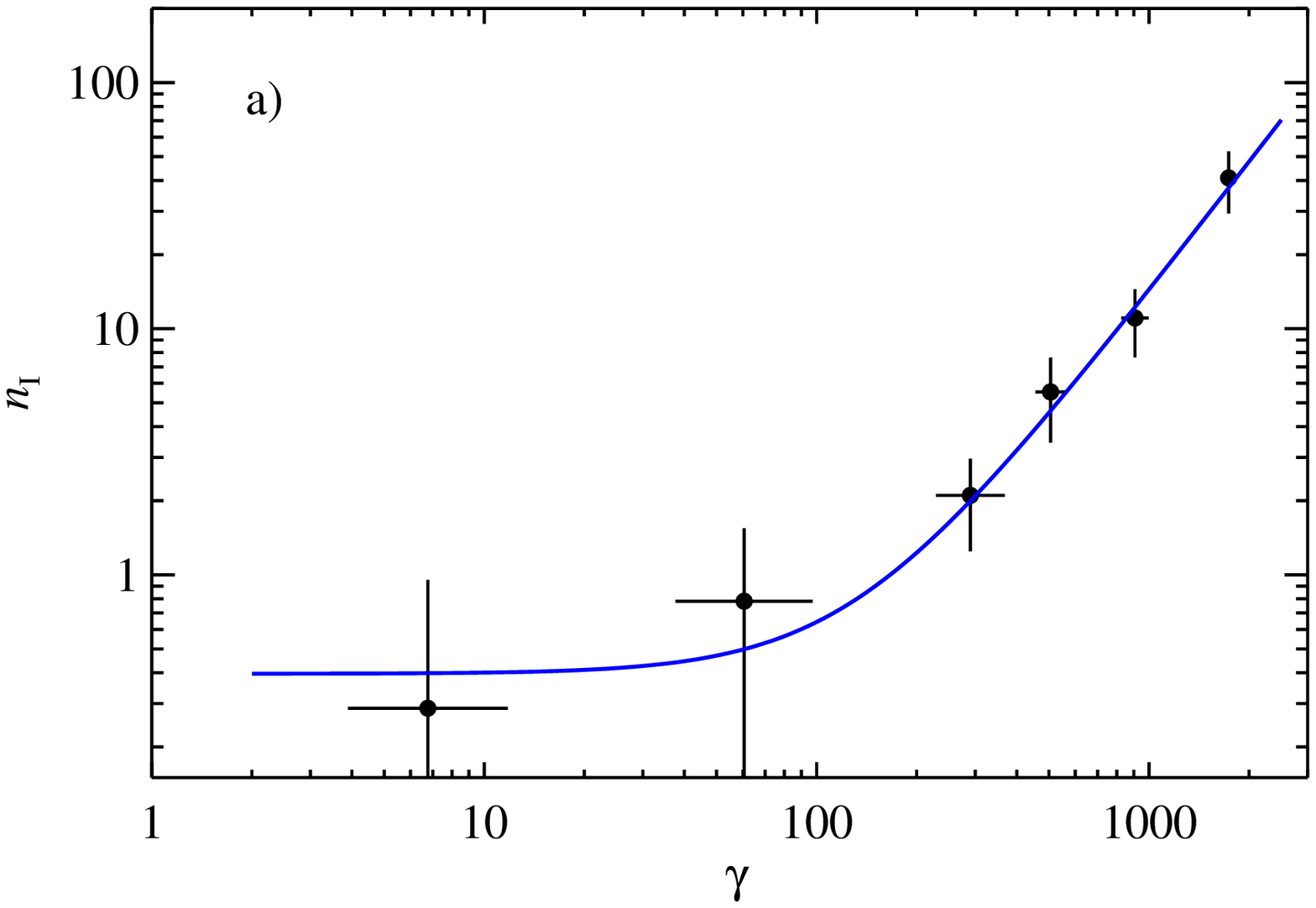}\hglue0.5cm                 
\includegraphics[width=0.47\textwidth]{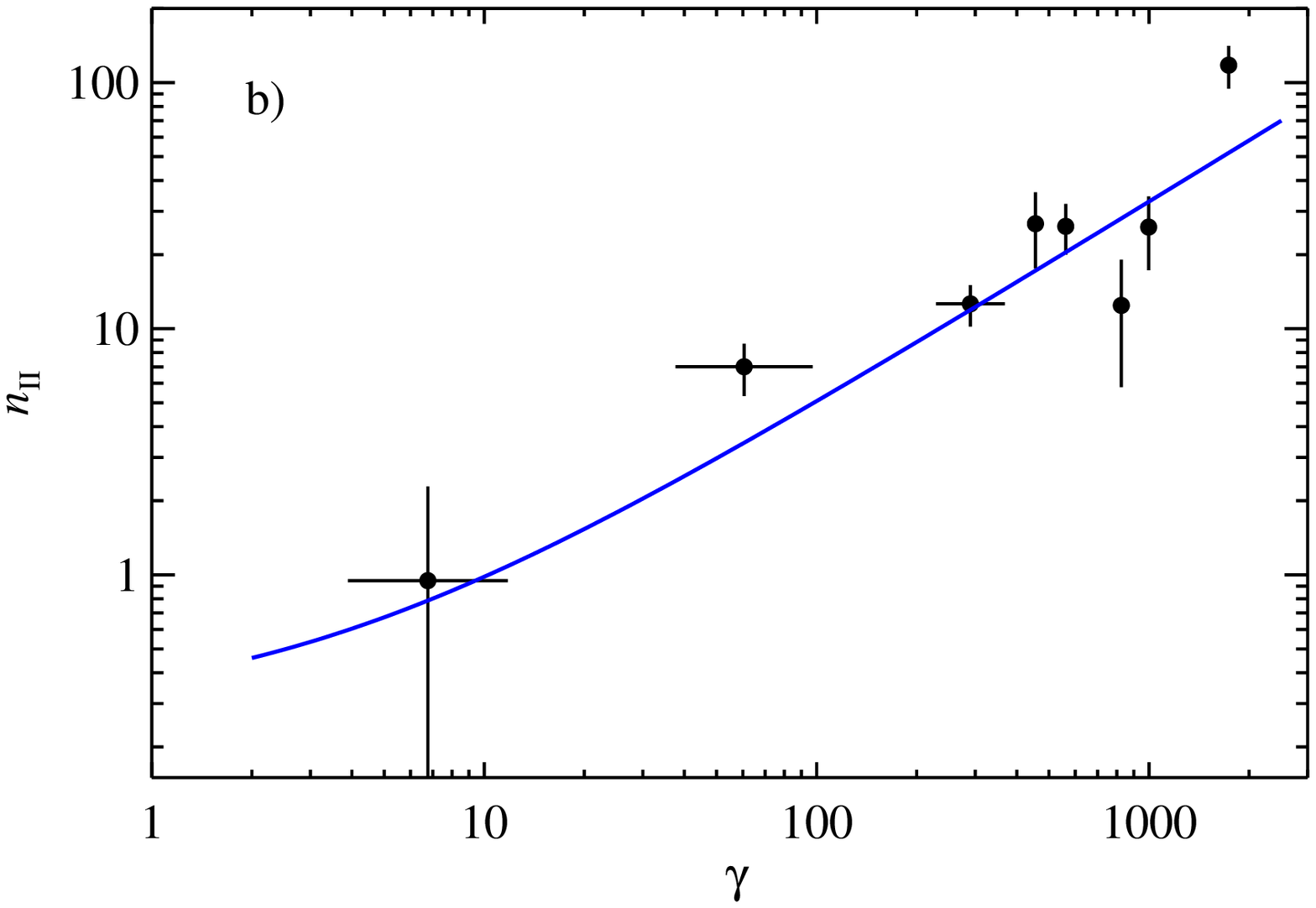}                         
\vglue0.3cm                                                                     
\includegraphics[width=0.47\textwidth]{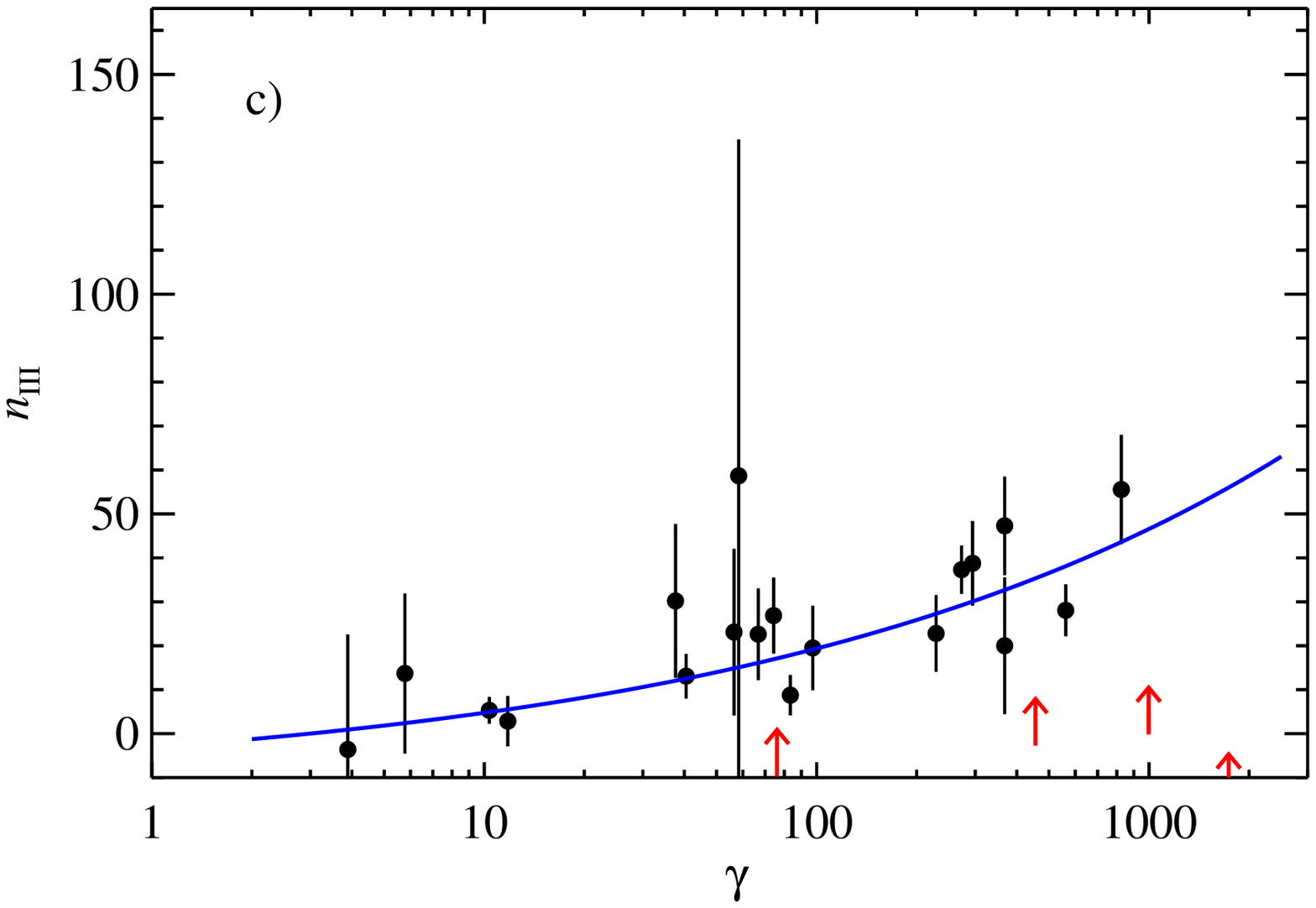}\hglue0.5cm           
\includegraphics[width=0.47\textwidth]{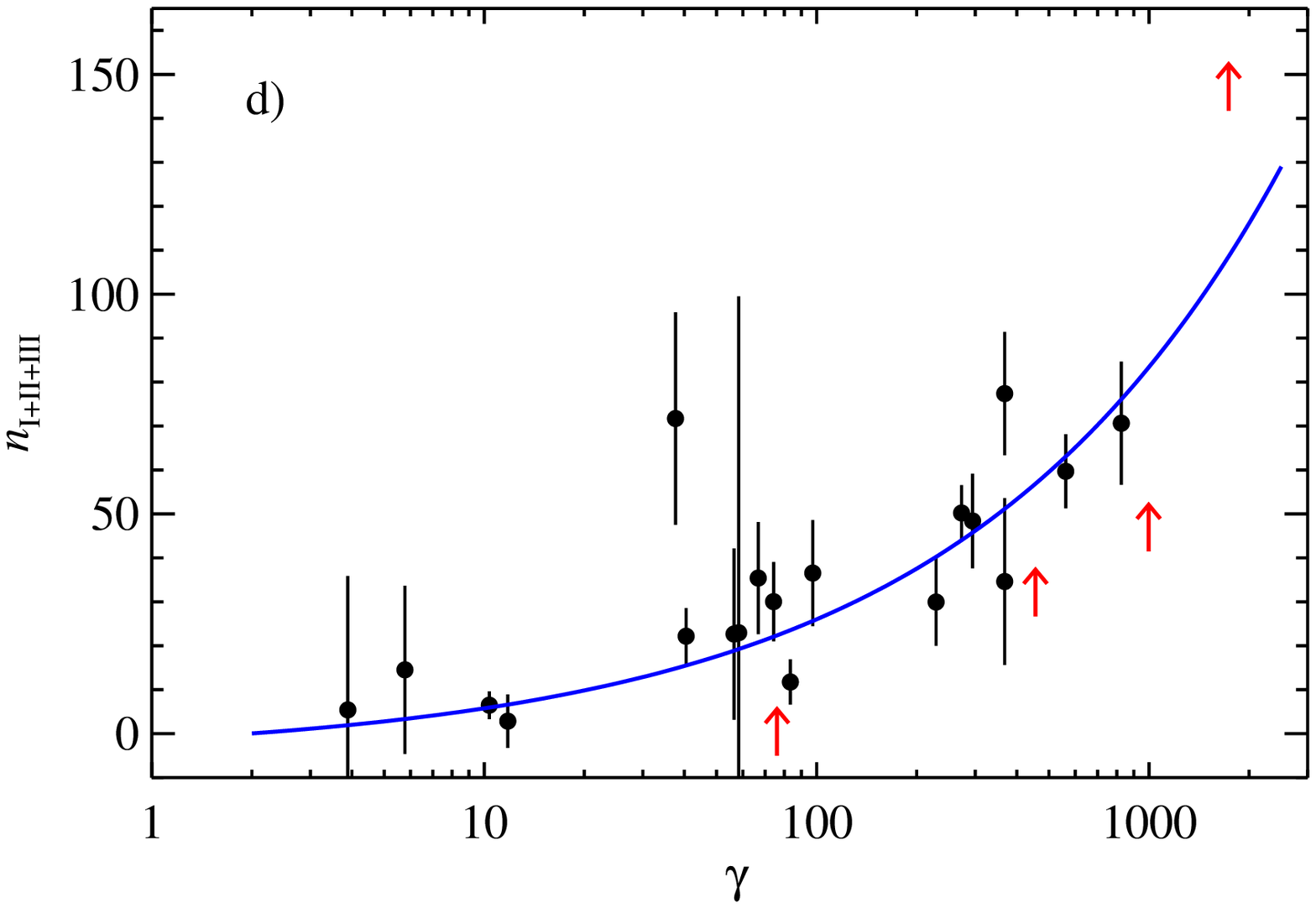}                           
\caption{The specific number of sources for each population ($n_x$) plotted as a function of the specific encounter frequency ($\gamma$), along with the best fit model of the form $C_x + A_x\gamma^{\alpha_x}$. a) Population I is almost exclusively quiescent LMXBs.  The data have been binned for the model fits and for plotting.  b) Population II is dominated by CVs.  The data were binned for plotting, but the fits were performed on the unbinned data.  c) Population III comprises many different source types.  The red arrows indicate clusters which were not observed deeply enough to complete the census of this population.  d) All sources above \ee{4}{30}~\ergsec\ are shown, similar to Fig.~2 from \citet{2003ApJ...591L.131P} but here in specific units (i.e., dividing all values by cluster mass).}
\label{fig:nvg}
\end{figure*}

For each population $x\in\left\{\mathrm{I, II, III, I+II+III}\right\}$ we plot $n_x$ as a function of $\Gamma$ (see Fig.~\ref{fig:nvg}).  We first fit a constant for each plot, which we find statistically unacceptable in all four cases.  The goodness of fit and the likelihood of each fit accurately describing the data (the $Q$ value, which is the probability of obtaining the observed $\chi^2$ assuming the model is correct and the best-fit parameters are the true parameters) are given in Table~\ref{tab:const}.  This can be interpreted as the likelihood of each of these populations being purely primordial.

As discussed above, we can then fit $n_x = C_x + A_x\gamma^{\alpha_x}$, with the two terms representing contributions from a primordial population (first term) and a dynamically influenced population (second term).  In Table~\ref{tab:powerlaw} we give the best fit parameters, the goodness of fit, and the probability that the model accurately describes the data.  We plot the best fit relationship in Fig.~\ref{fig:nvg}.

\begin{deluxetable}{ccccl}
\tablewidth{0pt}
\tablecaption{Results of fitting $n_x = C_x + A_x\gamma^{\alpha_x}$\label{tab:powerlaw}}
\tablehead{\colhead{$x$} & \colhead{$C_x$}  & \colhead{$\alpha_x$}& \colhead{$\chi^2/\mathrm{dof}$} & \colhead{$Q$ value}} \\
\startdata
I        & $0.40\err{0.53}{0.55}$ & $1.75\err{0.43}{0.36}$ & 0.45/3 & 0.93 \\
II       & $0.27\err{1.40}{1.99}$ & $0.83\err{0.29}{0.25}$ & 27.7/20 & 0.12 \\
III      & $-12\err{14}{235}$     & $0.27\err{0.24}{0.21}$ & 18.5/16 & 0.30\\
I+II+III & $-5.3\err{7.7}{15.5}$  & $0.45\err{0.17}{0.16}$ & 22.4/16 & 0.13
\enddata
\end{deluxetable}

The results for $n_\mathrm{I+II+II}$ are an extension of the work presented in \citet{2003ApJ...591L.131P}, including more clusters, and working now in specific units.  The results for the LMXB-dominated population ($n_\mathrm{I}$) are also an extension of the work presented in \citet{2003ApJ...591L.131P} and complementary to the work of \citet{2006ApJ...???..???H} who also consider the effects of metallicity as a determining factor in the number of LMXBs present in a cluster; they find no evidence for metallicity dependence.

The significant new result of this work is the demonstration that the CV-dominated population ($n_\mathrm{II}$) must also have a large dynamically-formed component.  A primordial-only population is clearly ruled out, and our simple prescription for primordial-plus-dynamical is found to be acceptable.  While our analysis has not taken into account a possibly confusing variation in IMF and other local primordial properties among globulars, such variations would not be expected to show the type of clear correlation between CV frequency and $\gamma$.

A fair question to ask is whether population II is really dominated by CVs.  This will only be resolved beyond a doubt when a significant majority of the sources in the population are identified.  The other likely major constituent is active main-sequence binaries.  As Fig.~\ref{fig:cmd} indicates, few of the identified main-sequence binaries in globular clusters are in this region, and, as noted above, very few active main-sequence binaries in the field reach luminosities above \ee{4}{31}~\ergsec.  We therefore feel justified in identifying population II as CV-dominated.  We do recognize that many of the non-CV sources in population II are binaries and, hence, can be expected to be overabundant because they also can be produced directly through encounters.

Note that our conclusion about the dynamical formation of CVs is totally independent of the uncertain abundance of CVs in the field.  It is therefore necessary to discuss the conclusion of \citet{2005ApJ...628..395T} that 47 Tuc shows no overabundance of CVs.  They present a prediction for the entire CV population, and here we deal with just the most X-ray luminous globular cluster CVs.  Assuming a roughly similar CV X-ray luminosity function in each cluster, this should not make a significant difference.  Note that the prediction of \citet{2005ApJ...628..395T} has a spread of a factor of three, and we are dealing here with numbers not much different than a factor of three.  However, we have overwhelming statistical evidence that the trend we find is real, even though the results based on any individual cluster alone would not be meaningful at all. 

It may be the case that the prediction of \citet{2005ApJ...628..395T} just happened to be in the right range of 47 Tuc.  For example, applying the prediction to $\omega$~Cen, which is more than twice as massive than 47 Tuc, would imply twice as many CVs.  In our CV-dominated population II, 47 Tuc has $\sim$16 sources whereas  $\omega$~Cen has only $\sim$2.   The comparison of \citet{2005ApJ...628..395T} with 47 Tuc, while plausible for a single cluster, does not
carry the statistical weight that we provide with our whole ensemble.

\section{Summary}
\label{sec:summary}
We conclude that in dense globular clusters, such as NGC 6388, NGC 6266 and 47 Tuc, the majority of CVs must be produced dynamically, by encounters
between single stars and/or binaries.  The obvious followup question is: what is the mechanism responsible?  That will need to be sorted out in detailed simulations, in which each cluster is modeled separately.  This will be done, but it will take several years.  However, for now, we can already draw the firm conclusion that these CVs are predominantly formed dynamically by ruling out a primordial-only scenario, which has only a \ee{1.97}{-9} probability of describing the observations (Table~\ref{tab:const}).

Because of the long-standing theoretical predictions that globulars should be rich in white dwarf binaries, they have been called Type Ia supernova factories \citep{2002ApJ...571..830S}.  We have now established observational evidence to support this idea.  Globular clusters are adept at forming exotic binaries and their offspring: neutron star binaries, millisecond pulsars, and, as we have now shown, white dwarf binaries.  For this reason, globular clusters may play an interesting and significant role in the story of the short/hard gamma-ray bursts \citep{2006ApJ...???..??Z} and provide exciting targets for gravitational wave detectors \citep{1999ApJ...520..233B}.

\end{document}